\documentclass[journal]{IEEEtran}
\usepackage{cite}
\usepackage{amsmath,amssymb,amsfonts}
\usepackage[normalem]{ulem}
\usepackage{graphicx}
\usepackage{textcomp}
\usepackage{siunitx}
\usepackage{algorithm}
\usepackage{algpseudocode}
\usepackage{fancyhdr}
\usepackage{xcolor}
\def\BibTeX{{\rm B\kern-.05em{\sc i\kern-.025em b}\kern-.08em
    T\kern-.1667em\lower.7ex\hbox{E}\kern-.125emX}}

\fancypagestyle{cfooter}{ %
\fancyhf{} 

}

\begin{document}

\title{FRAP analysis\\ Measuring biophysical kinetic parameters using image analysis}

\author{Sharva V Hiremath, North Carolina State University\\
        Dr. Etika Goyal, Texas AM University\\
        Dr. Gregory Reeves, Texas AM University\\
        Dr. Cranos Williams, North Carolina State University}


\author{\IEEEauthorblockN{Sharva V Hiremath\IEEEauthorrefmark{1} \, \, \, Etika Goyal\IEEEauthorrefmark{2} \, \, \,Gregory Reeves\IEEEauthorrefmark{2} \, \, \,Cranos M. Williams\IEEEauthorrefmark{1}\\}
\IEEEauthorblockA{\IEEEauthorrefmark{1}\textit{Department of Electrical and Computer Engineering, North Carolina State University}, Raleigh, U.S.A. \\
\IEEEauthorrefmark{2}\textit{Department of Chemical Engineering, Texas A\&M University}, College Station, U.S.A.}}



\maketitle

\begin{abstract}
Understanding transcription factor dynamics is crucial for unraveling the regulatory mechanisms of gene expression that underpin cellular function and development. Measurements of transcription factor subcellular movements are essential for developing predictive models of gene expression. However, obtaining these quantitative measurements poses significant challenges due to the inherent variability of biological data and the need for high precision in tracking the movement and interaction of molecules. Our computational pipeline provides a solution to these challenges, offering a comprehensive approach to the quantitative analysis of transcription factor dynamics. Our pipeline integrates advanced image segmentation to accurately delineate individual nuclei, precise nucleus tracking to monitor changes over time, and detailed intensity extraction to measure fluorescence as a proxy for transcription factor activity. Combining our pipeline with techniques such as fluorescence recovery after photobleaching enables the estimation of vital biophysical parameters, such as transcription factor import and export rates.
\end{abstract}


\section{Introduction}

Regulation of gene expression is critical across all areas of biology, often mediated by transcription factors that enter the nucleus to bind to specific DNA sequences, thereby controlling gene activation or repression. For a comprehensive understanding of the relationship between transcription factor dynamics and gene expression, it is imperative to quantitatively analyze the movement of transcription factors within cells \cite{mir2018dynamic,de2022following,givre2023modulation,lu2021transcription}. Specifically, insights into the rates of nuclear import and export of transcription factors are vital, as they elucidate how variations in the bulk concentration of transcription factors within the nucleus can influence gene expression decisions, thereby affecting cellular function and organismal development \cite{gregor2007stability, clark2016tracking, carrell2017facilitated, al2018dorsal}. 

In this context, our paper leverages these insights to quantify the nuclear import and export rates of the transcription factor Dorsal (Dl), which exhibits a spatial gradient of nuclear concentration in 1-3 hour old \textit{Drosophila melanogaster} embryos \cite{schloop2020formation}. Quantifying the dynamics of transcription factors like Dorsal presents several challenges, including the inherent variability of biological data over time and the need for precise tracking and analysis of molecular movements. Additionally, accurately capturing these dynamics requires computational tools and imaging techniques that can handle the rigors of live cell imaging without disrupting cellular processes. Our approach effectively extracts biophysical parameters related to Dorsal, demonstrating the efficacy of our pipeline.

Fluorescence Recovery After Photobleaching (FRAP) assays are a critical tool for quantifying protein movement. Introduced in the 1970s primarily to observe the mobility of fluorescently tagged molecules on cell surfaces \cite{axelrod1976mobility}, FRAP assays have evolved significantly with the advent of Green Fluorescent Protein (GFP) technology. This evolution has broadened their application in exploring protein behaviors within cells \cite{houtsmuller2005fluorescence,loren2015fluorescence}.

FRAP assays provide a distinctive perspective on protein dynamics. By intentionally bleaching a specific area in a cell or tissue and observing how fluorescence returns, researchers can extract valuable information about protein movement, interaction, and turnover rates \cite{ishikawa2012advanced,sprague2005frap,mueller2010frap,carrero2003using}. The strength of FRAP assays lies not just in their capacity to yield quantitative data, but also in their suitability for live cell and tissue studies, rendering them crucial for investigating protein dynamics. However, computer vision and image analysis pipelines are essential to derive biophysical parameters from fluorescent image data. These tools are crucial for processing the visual data, enabling the extraction and quantification of vital biological information embedded within the images. By leveraging image processing algorithms and computational methods, these pipelines transform raw image data into insights, which advance our understanding of cellular processes.

To overcome the challenges associated with data variability and the requirement for attributes such as intensity or shape to remain consistent across different locations and time frames, we have developed a new computational pipeline. This pipeline merges FRAP assays with image analysis techniques to derive quantitative measurements. Unlike the numerous existing methods proposed for locating/segmenting nuclei and tracking them \cite{irshad2013methods,hayakawa2021computational,beevi2016detection,song2017dual,shi2016automated,neghina2016automatic,al2009improved,qi2013drosophila,amat2014fast,veta2011marker,ahasan2016white}, which we have dismissed due to these challenges. Our approach is designed to address the challenges of accurately segmenting irregularly shaped nuclei, tracking nuclei through different time frames in the presence of movement, and estimating changes in fluorescence intensity in nuclei and surrounding cytoplasm. 

The image analysis pipeline consists of three major steps: (1) image segmentation, to accurately delineate individual nuclei and their cytoplasmic compartments from the surrounding biological milieu; (2) tracking, to monitor the movement and changes in the position of each nucleus (and their surrounding cytoplasm) over time; and (3) intensity extraction, to measure the fluorescence intensity as a proxy for various biological phenomena. Each of these steps is designed to ensure the integrity of the data is preserved, facilitating a seamless transition to the data fitting phase. Here, the extracted parameters serve as the foundation for in-depth analysis and interpretation as they provide empirical evidence that can be used to test hypotheses about cellular processes, particularly the dynamics of transcription factors like the import and export rates in and out of the nucleus. By quantifying these dynamics, researchers can better understand how transcription factors regulate gene expression, how they respond to various stimuli, or how mutations might affect their function. Moreover, the parameters allow for the creation of detailed models of cellular behavior and predict future behaviors and responses under various conditions. 

This pipeline is advantageous in scenarios where traditional methods prove inadequate due to the challenges associated with the data. Furthermore, our approach segments, tracks, and measures the intensities of the surrounding cytoplasm as well (not just the nuclei), which helps improve the quantification of the dynamics of transcription factors by providing additional information for the fitting phase. Through the application of our pipeline, researchers are equipped to navigate the complexities of biological data, enabling a deeper understanding of the underlying mechanisms of the transcription factors at play.

\section{Materials and methods}

\subsection{Data collection}
\begin{figure*}[!t]
\centering
{\includegraphics{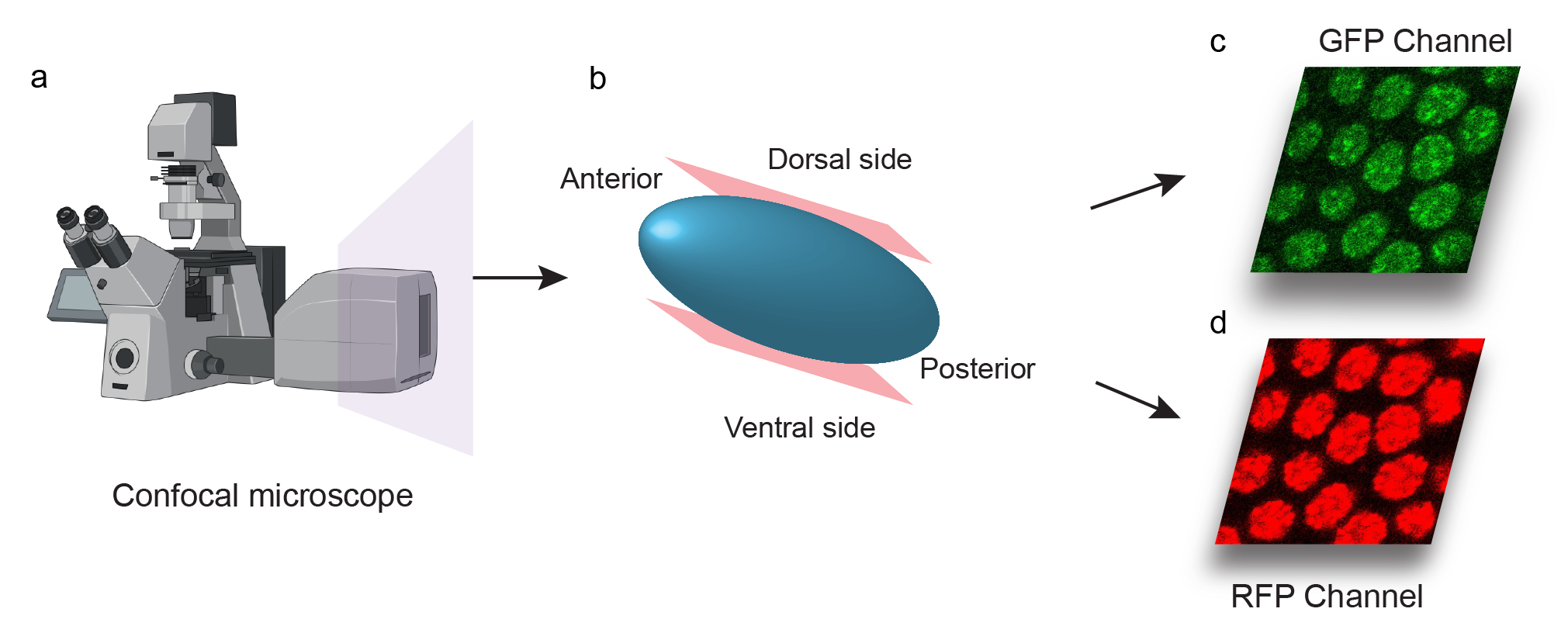}}%
\caption{FRAP experiment setup and data collection. (a) Illustration of the setup for collecting the image data from FRAP experiments using a confocal microscope (\textit{Created with BioRender.com}). (b) The data is collected from 2 cross sections of the embryo, i.e., the dorsal and ventral side, for measuring import/export rates. (c,d) Image data collected for 2 channels.}
\label{fig:FRAP setup}
\end{figure*}

\begin{figure*}[!t]
\centering
{\includegraphics{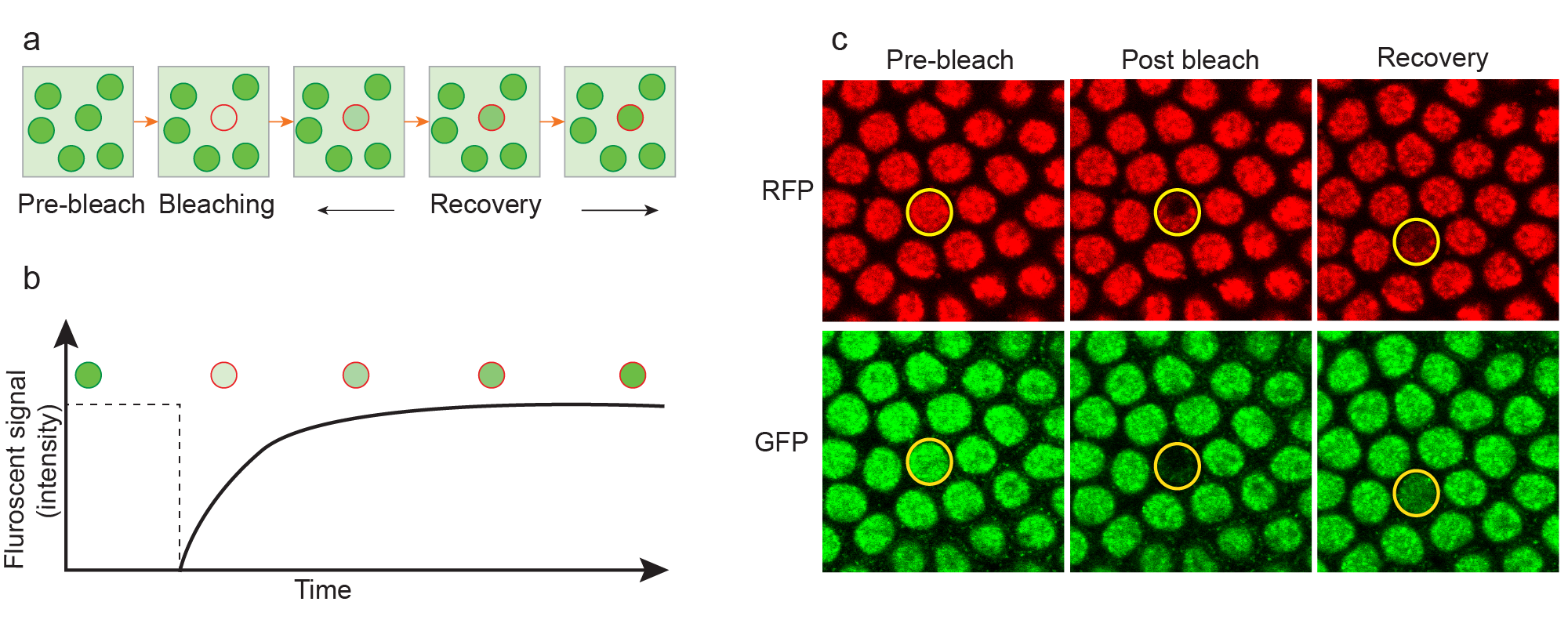}}%
\caption{(a) In FRAP assays, the time that fluorescent molecules need to replenish a bleached-out area is measured, basically assessing how quickly a dark sample area turns bright again \cite{blassle2018quantitative}. We photobleach one nucleus using a high-power laser pulse to alter the fluorophore molecules attached to the protein, such that they are unable to fluoresce. (b) Illustration of the recovery of fluorescence observed in the bleached nucleus over time. The bleached nucleus (highlighted by a red circle) undergoes an exchange of protein tagged with FP's from the surrounding cytoplasm. This exchange with the surrounding cytoplasm leads to the eventual recovery in intensity observed. (c) Here we have an example of what the data looks like. The bleached nucleus is indicated by the circle in yellow. We can observe how the intensity of the bleached nucleus disappears initially after bleaching before reappearing later in GFP but remains visible throughout in RFP.}
\label{fig:FRAP assays}
\end{figure*}

The data collected to measure the import/export rates uses two fluorescent proteins (FPs) captured across two channels, which collect structural information and protein signal data (Fig. \ref{fig:FRAP setup}). The first channel (the nuclear channel) captures the structural information where the nuclei are highlighted as histone is attached with a red fluorescent protein (RFP). The second channel (the nuclear protein channel) is one where we actually observe the change in intensity of the bleached nucleus by tagging the gene of interest, in this case Dorsal (Dl), with a green fluorescent protein (GFP). The data are collected from two locations of the embryo (dorsal and ventral), each looking at the behavior of the tagged protein at different locations along the DV-axis. (Fig. \ref{fig:FRAP assays}).


A single nucleus \cite{carrell2017facilitated} was bleached at 100\% laser power. We measured the recovery of the nuclear fluorescence, achieved by the movement of cytoplasmic Dl-GFP molecules into the bleached nucleus to replenish the bleached-out nucleus. This recovery quantifies the rate at which the protein (by proxy of relative intensity) moves in and out of the nucleus (Fig. \ref{fig:FRAP assays}). 

We use the nuclear channel to locate/segment the nuclei. Next, we map this location information on the protein channel to derive intensity measurements of the bleached nucleus, tracking changes from the moment of bleaching up to 30 minutes afterward. Finally, a model of nuclear import and export is fitted to these values to estimate the import/export rates.

\begin{figure}[!t]
\centering
\includegraphics[width=2.5in]{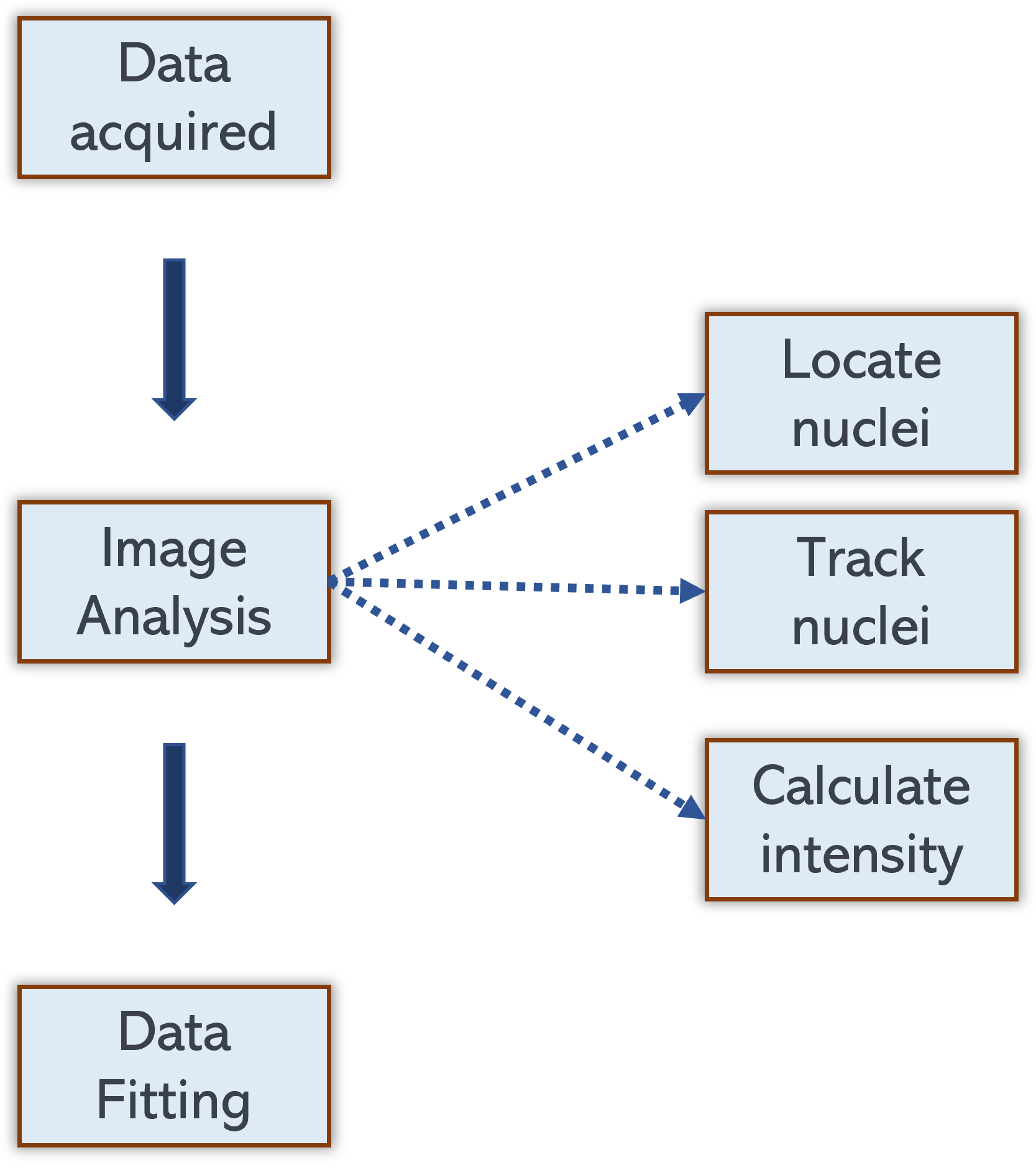}
\caption{Broad overview of steps in the pipeline for analysing FRAP images}
\label{fig:FRAP pipeline}
\end{figure}

\subsection{Challenges with image data}

The methodology for analyzing microscopy images is intricately linked to the complexities of the data obtained from experimental processes. These challenges, which significantly influence the analysis approach, are multifaceted and cover various issues related to the dynamic and variable characteristics of the observed specimens.

One primary challenge involves intensity variation, where the brightness of individual nuclei changes over time, coupled with differences in intensity observed among nuclei within the same time frame. This variability complicates the consistent observation and analysis of these nuclei. 

In addition to changes in internal structure, the shape of nuclei can vary over time and across different nuclei within a single observation period. These shape changes, characterized by unique and irregular contours, pose additional challenges in tracking and analyzing each nucleus accurately. The proximity of nuclei to one another further complicates this task, as it becomes challenging to distinguish between closely situated nuclei, to ascertain individual identities.

High levels of movement among the nuclei introduce further difficulties in tracking and maintaining consistent identification. This is particularly problematic when nuclei exhibit rapid or extensive motion, which can lead to errors in tracking continuity. Lastly, the issue of target re-identification emerges when nuclei move in and out of the observational frame, complicating the task of keeping consistent labels on them throughout the study period. This issue occurs because the observed region under the microscope is a small area (32$\mu$m x 32$\mu$m) of the embryo, allowing nuclei near the edges to exit and re-enter the frame during the observation period.


Given that intensity measurements are used as proxies for the relative concentration of Dl in nuclei and cytoplasm, which is crucial for assessing import and export rates, issues such as changes in nuclear shape, irregular contours, and movement of nuclei present significant challenges. These complexities demand analytical approaches and methodologies to accurately interpret microscopy images and derive meaningful insights from experimental data.

\begin{figure}[!t]
\centering
\includegraphics{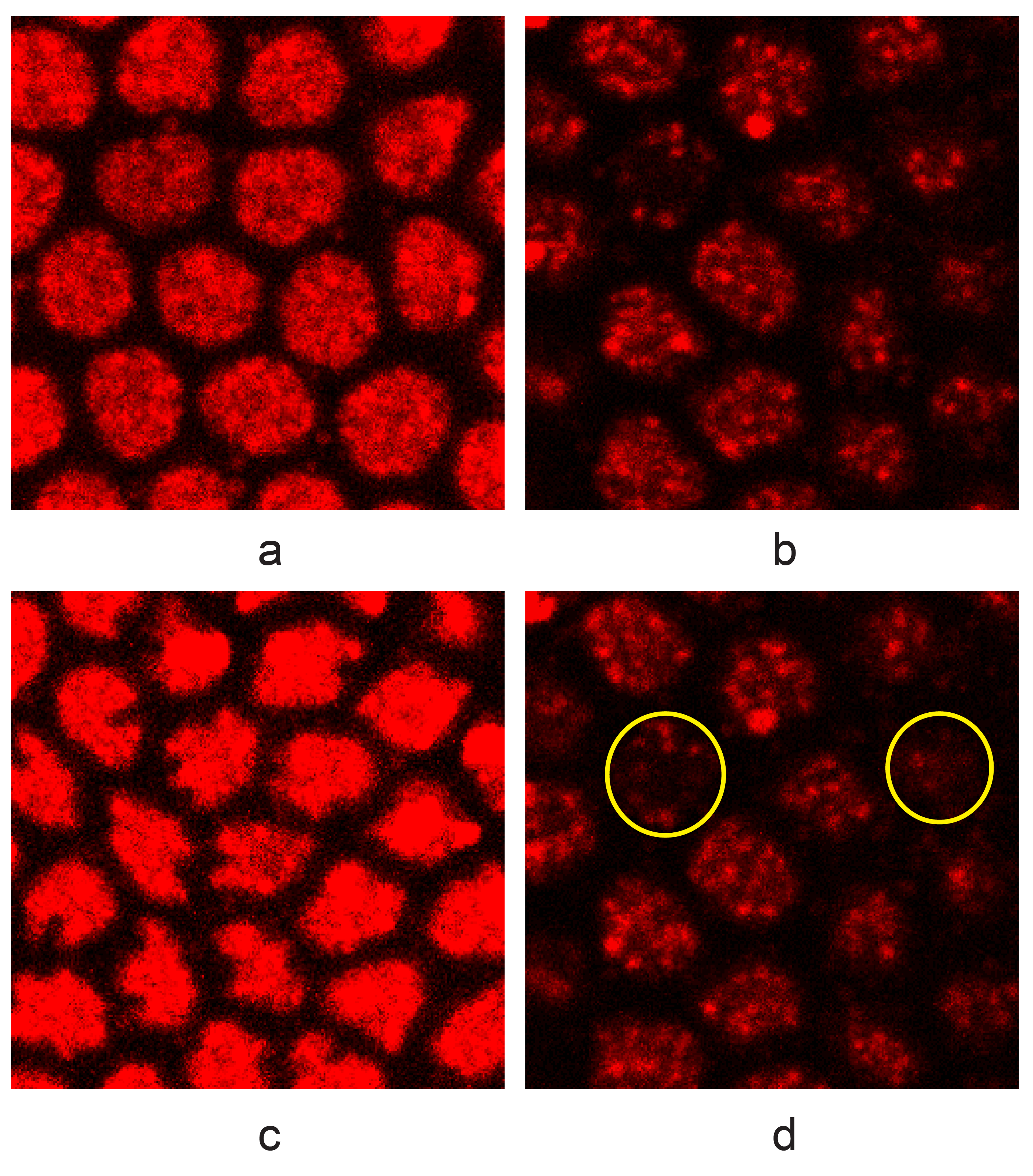}
\caption{Challenges in nuclei detection. (A) Idea case for nuclei (B) Variation in intensity within nuclei (C) Irregularly shaped nuclei (D) Nuclei breaking into blobs are encircled
}
\label{fig:frap_data_challenges}
\end{figure}

\subsection{Proposed approach overview}

The analysis of microscopy data collected from FRAP experiments can broadly be classified into two parts, namely image analysis and data fitting. Image analysis consists of locating the nuclei, tracking the bleached nucleus and surrounding nuclei, and extracting the intensity of nuclei from the protein channel. Data fitting consists of empirically fitting these intensities to a differential equation to obtain the import/export rates (Fig. \ref{fig:FRAP pipeline}). 

Tracking the bleached nucleus through the different time frames is essential as we need to measure how the fluorescence recovers over time in the bleached nucleus. In addition, due to unintentional bleaching, in which the regular acquisition of data slowly bleaches all GFP molecules in the frame, it is equally important to track the other surrounding nuclei as well. If the intensity of the surrounding nuclei decreases over time, it can be attributed to unintentional bleaching. We use the measurements of intensities of the surrounding nuclei to normalize the recovery of the intentionally bleached nucleus (see Extracting intensities of nuclei and cytoplasm).

For analyzing FRAP images and solving the challenges associated with it, we used watershed for segmentation \cite{meyer1994topographic}. It allows us to deal with local issues individually instead of resorting to a global-based solution. This approach suits us well as it simultaneously addresses the variations in intensity, shape, and overlapping nuclei.

After segmenting the nuclei, we initiate the tracking process. Our approach is influenced by several constraints, including the potential for high movement (due to data being collected close to gastrulation) and the movement of nuclei in and out of the frame. These factors introduce complexities in simultaneously tracking multiple nuclei across frames exhibiting changing intensity profiles, evolving shapes, and nuclei entering or exiting the frame. In response to these challenges, we employ a point registration approach \cite{emami2021computerized,mount1998improved,goshtasby2012image,yuan20163d}.
This method allows for the simultaneous tracking of multiple objects, obviating the need to search each frame for matching objects in a sequential manner, which increases the overall complexity.

\subsection{Segmentation}
We perform pre-processing to prepare the data for segmentation. First, we correct image defects caused by uneven illumination in fluorescent microscopy using background correction by multiplying the image data (array) with a scalar value (determined heuristically). This step helps improve the illumination of both nuclei and cytoplasm, making them sufficiently visible to be accurately identified during the segmentation process. We use Gaussian blurring to smoothen the disparities in intensity within nuclei caused by an accumulation of fluorophores (FPs) in a few locations. Incorporating Gaussian blurring in the preprocessing phase helps to mitigate the issue of oversegmentation, since it improves the determination and selection of appropriate nuclear and background markers.

\begin{figure*}[!t]
\centering
\includegraphics{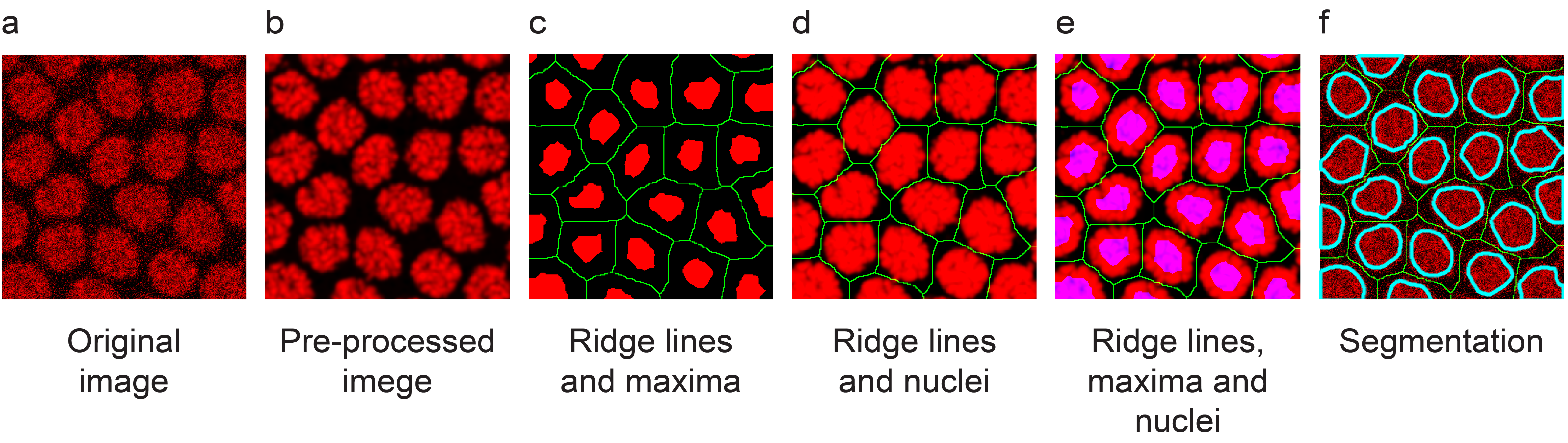}
\caption{Different steps of the Image analysis pipeline for FRAP. (A) Raw image data (B) Image after preprocessing using background correction, blurring, and removing small artifacts  (C) Regional maxima (in red) and the ridge lines (in green) obtained using watershed (D) Processed data (in red) and ridge lines (in green) (E) Processed data (in red), ridge lines (in green) and regional maxima (in pink) (F) original data, ridge lines (in green) and the nuclear masks obtained using Otsu's thresholding (in cyan)}
\label{fig:frap_sample_pipeline_results}
\end{figure*}

Next, we calculate the regional maxima for the images, and these maxima are used as foreground markers for the marker-based watershed. Using the watershed technique allows us to obtain ridge lines (watershed lines) that separate the image into compartments that each contain a single nucleus. This is effective as we can now split each nucleus into its own territory and process it without affecting the others. Using this approach also alleviates issues where global approaches to segmenting nuclei do not work due to variations in intensity among the nuclei and other local issues. Additionally, we retain these ridge lines because they help link cytoplasmic regions (areas in the cell outside the nucleus but within the cell membrane) to various nuclei. These associations are utilized later in the pipeline to measure the intensity of the cytoplasm surrounding both the bleached nucleus and other nearby nuclei.

After isolating the individual compartments, we create a set of temporary, empty arrays, ensuring the count of these arrays corresponds exactly to the number of compartments. Each array is created to match the original image size, and these arrays are populated with the pixel intensities from the nuclear channel that correspond to each specific compartment.
Following the transfer of pixel intensities, we apply Otsu’s method \cite{otsu1979threshold} for thresholding, which produces a binary mask for each compartment. This approach is effective since we are distinguishing between two primary classes: the nuclei and the cytoplasm. After generating the binary masks for each individual compartment in the temporary arrays, we then stitch together all the binary masks obtained for each compartment containing the nuclei. This merging process yields a unified mask that encompasses all the nuclei across the different compartments.

Once a complete binary mask is determined, it is then cleaned morphologically using erosion followed by dilation with disk-shaped structuring elements of sizes 5 pixels ($\sim$ .31 $\mu$m) and 1 pixel ($\sim$ .062 $\mu$m), respectively. This procedure corrects any irregular boundaries and produces a conservative mask, which is particularly useful for the nuclei that might be overlapping. Additionally, we implement a size-based filter to exclude artifacts smaller than 400 pixels. This exclusion criterion is based on the analysis of nuclear size histograms, which indicate an average nuclear radius of approximately 40 pixels ($\sim$ 2.5 $\mu$m), and an average nuclear area of 5026 pixels. Notably, nuclei that are only partially within the frame exhibit much smaller pixel areas. Removing significantly smaller artifacts prevents the accidental inclusion of cytoplasmic components, ensuring the mask accurately represents nuclear regions only.

This cleaned binary mask is then used to create a nuclear mask that outlines the individual nuclei border. In addition to the nuclear mask, details such as average intensity, pixel locations, and centroid locations are also stored. We can actively utilize these details to analyze the trajectory of nuclei as required.

\subsection{Tracking}

Once we have the segmented nuclei for each time frame, we track each nucleus over time. We used a points registration algorithm in which the centroid coordinates of each nucleus, as found by the nuclear segmentation, are globally correlated between frames (Fig. 6). Thus, the persistence of labels from one frame to the next were the result of an optimization algorithm in which each nucleus was assigned the label to the closest nucleus from the previous frame.

For the first frame, we label the nuclei. In the following frames, we use the locations of the centroids (stored as point clouds) of nuclei detected to identify the nuclei from the previous frames. Defining the constellation of centroids at time 't' as '$C_{t}$' we identify the nuclei by comparing the constellation of centroids '$C_{t-1}$' from frame 't-1' with '$C_{t}$' by registering them with each other using an Iterative Closest Point Algorithm \cite{Chen1992ObjectMB}.

Registering the point clouds gives the transformation matrix which, when applied on '$C_{t}$', gives a new set of points '$C'_{t}$' that best match with '$C_{t-1}$'. Using this, we can infer the identity/label of the nuclei based on which nucleus is closest to it in the transformed coordinates. 


In our dataset, the nuclei exhibit coordinated motion, forming a constellation. This pattern means that while the nuclei move, they are bounded by other nuclei within the observed frame, limiting their freedom to move freely. As a result, their relative movement is constrained, which assists in tracking them more effectively. We can trace the nuclei by relying on their relative positions, thereby addressing the challenge of tracking during periods of significant movement and ensuring accurate identification throughout the observation period.

\begin{figure}[!t]
\centering
\includegraphics{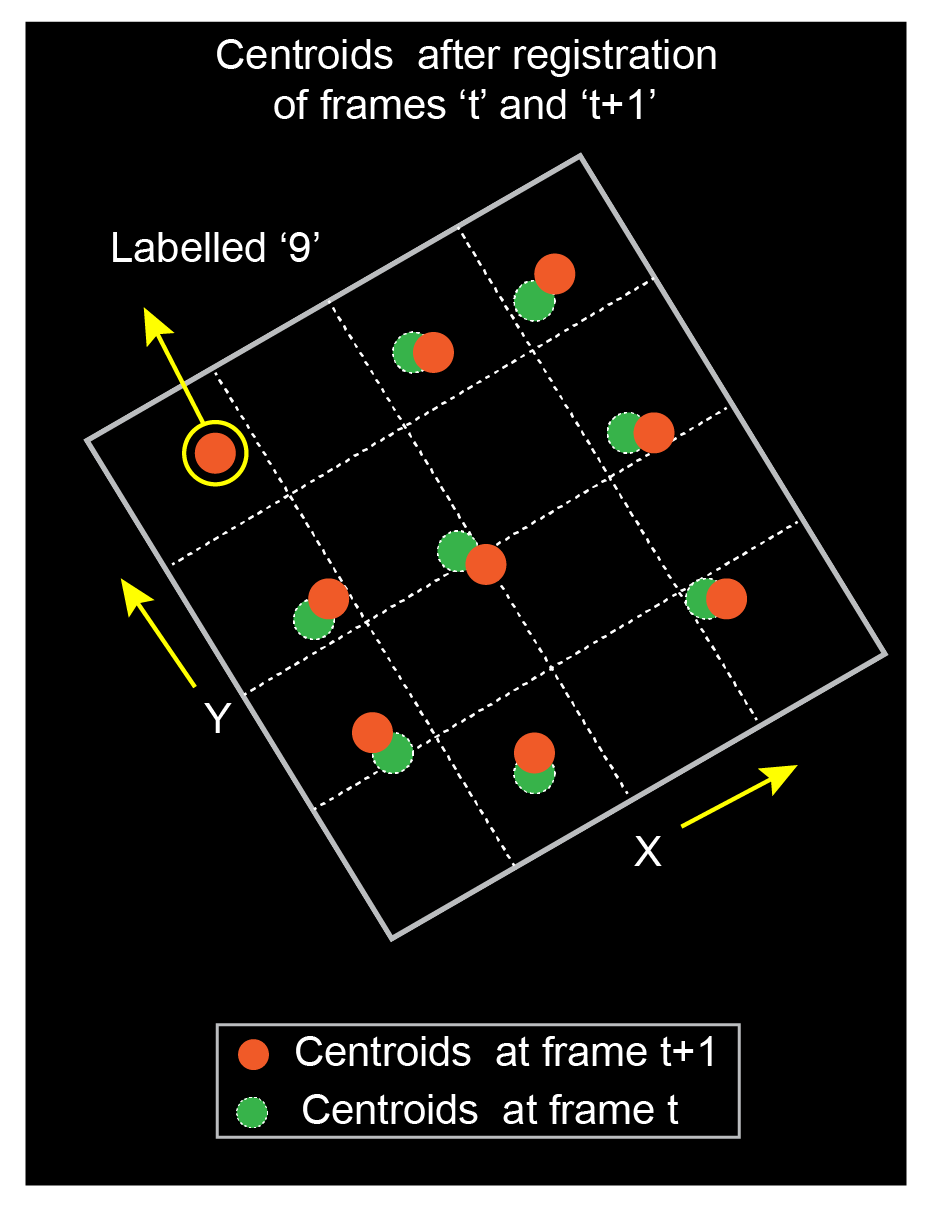}
\caption{Example illustrating tracking using Point registration approach. Here we have centroids from frame 't' and the transformed centroids from frame 't+1' juxtaposed on top of each other. We assign labels to the centroids from 't+1' based on analysing the closest centroid from frame 't'. If some centroids haven’t appeared previously, we assign them new labels, like the centroid labelled ’9’, which was not present earlier.}
\label{fig:frap_tracking_example}
\end{figure}

To solve the other problem of target re-identification, we add a "lookback feature" in the algorithm. This lookback feature allows for examining previous frames to check for matching points if there are points in '$C'_{t}$' that do not correspond with any points in '$C_{t-1}$'. The number of frames subject to retrospective examination is defined by a user-specified parameter (default set at 3). If a nucleus is missing for a number of frames larger than the lookback parameter, then the nucleus is assigned a new label in case it re-emerges. Consequently, should matching points be identified, the nearest label is assigned under the presumption of a nucleus reappearing. In the absence of corresponding points, these unmatched points are classified as new nuclei and are accordingly assigned a new label, while ensuring conservation of label number.

\paragraph{Missing bleached nucleus}

\hfill \break
In some instances, following the bleaching event, the bleached nucleus disappears from both the nuclear and protein channels. To address this scenario, we employ the positional data of neighboring nuclei to trace the location of the bleached nucleus. Our approach hinges on the assumption that the bleached nucleus maintains its shape throughout the frames (3-4 frames) during its absence. Leveraging this assumption, we project the nuclear mask of the bleached nucleus across the missing frames by computing the average movement of the adjacent nuclei.

\subsection{Extracting intensities of nuclei and cytoplasm}

We apply the masks derived from segmentation and tracking to the protein channel to calculate the mean intensities. For each nuclear and cytoplasmic compartment, we compute the mean pixel intensity, excluding the lowest and highest five percentiles to mitigate the influence of outliers and noise on the data. This ensures the calculated means accurately reflect the true intensity distribution. Subsequently, we adjust these mean values by subtracting the background intensity.

After extracting the intensities for all the nuclei and cytoplasmic compartments for all the time frames, we examine if any unintentional bleaching has occurred. This issue arises when all the GFP molecules (outside of the bleached nucleus) in the frame get bleached over time during data collection. We detect if unintentional bleaching has occurred by looking at the intensities of the neighboring nuclei and cytoplasm. The intensity measurements of the neighboring nuclei and cytoplasm are used to normalize the intensity of the bleached nucleus and its cytoplasmic compartment to accommodate the effects of intentional bleaching. The centroids of the nuclei were utilized to ascertain the proximity of neighboring nuclei. By calculating distances between the bleached nucleus and others, it is possible to identify which nuclei are nearest to the bleached nucleus. A user-defined parameter determines the specific count of nuclei to be regarded as neighbors in this context.

We employ Equation (\ref{intensity norm}) to normalize the intensities of the bleached nucleus and its cytoplasm. Here, '$y_{norm}$' represents the normalized intensity for either the bleached nucleus or cytoplasm, '$y_{blc}$'denotes the intensity of the bleached nucleus or cytoplasm, '$y_{mean}$' indicates the average intensity of surrounding nuclei or cytoplasmic regions, and '$y_{mean}^{0}$' refers to the initial average intensity of these surrounding areas immediately after the bleaching event.

\begin{equation}\label{intensity norm}
y_{norm} = \frac{y_{blc}}{y_{mean}}*y_{mean}^{0}
\end{equation}

\subsection{Estimating import and export rates using fitting}

Using the extracted intensities of the bleached nucleus ('$y_{nuc}$') and its cytoplasm '$y_{cyt}$' in nuclear protein channel we can estimate the import and export rates. 

Equation (\ref{nuc conc of C}) \cite{carrell2017facilitated} is used to describe the nuclear concentration of the protein (Dl), where  $[Dl]$ is the concentration of the protein, $k_{in}$ is the import rate and $k_{out}$ is the export rate.

\begin{equation}\label{nuc conc of C}
\frac{d[Dl]_{nuc}}{dt}=k_{in }[Dl]_{cyt} - k_{out}[Dl]_{nuc}
\end{equation}

Equation (\ref{fit eqn for kike}) is the solution for the differential equation,

\begin{equation}\label{fit eqn for kike}
\begin{aligned}
\relax [Dl]_{nuc}(t)=c_0 \exp(-k_{\text {out }} t)+\\k_{in} \exp (-k_{out } t) \int_0^t[Dl]_{cyt }(t^{\prime}) \exp(k_{\text {out }} t^{\prime}) d t^{\prime}
\end{aligned}
\end{equation}

 The time-course measurements of cytoplasmic intensities ($y_{cyt}$) serve as inputs to equation (\ref{fit eqn for kike}) for calculating the fitted nuclear protein concentrations ($y^{fit}_{nuc}$). We use non-linear least squares fitting to find $k_{in}$ (import rate) and $k_{out}$ (export rate) such that the difference between the fitted intensities and measured intensities is minimized.

This approach facilitates the precise estimation of import and export rates for the targeted protein based on the FRAP assay data available.

\section{Results and discussion}

\begin{figure*}[!t]
\centering
\includegraphics{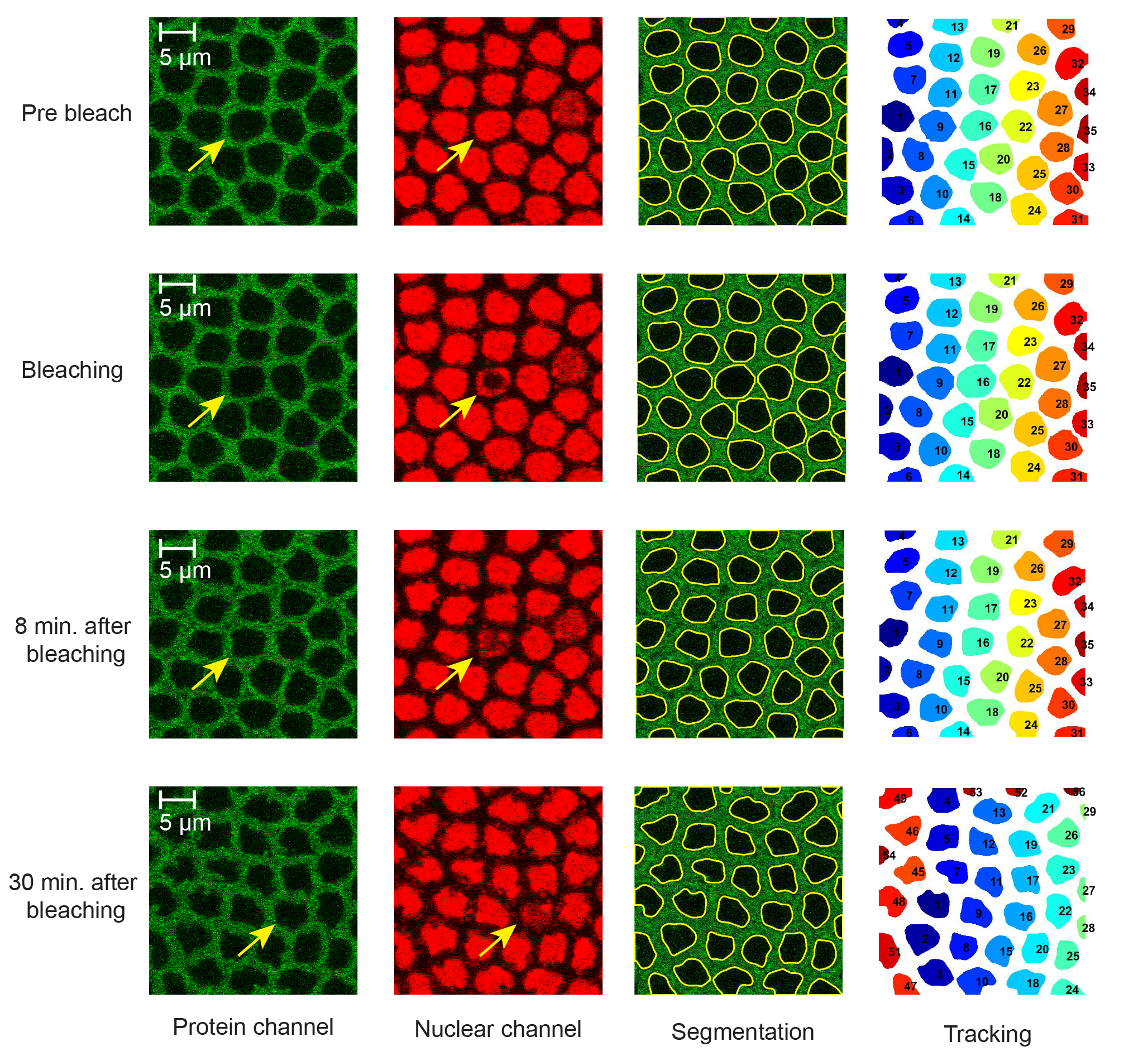}%
\caption{
Results of image analysis when the nuclear int. $<<$ cytoplasmic intensity. The bleached nucleus (label number 16) was segmented and tracked through all the time frames along with its neighbors. }
\label{fig: image analysis nuc low cyt high}
\end{figure*}
\begin{figure*}[!t]
\centering
\includegraphics{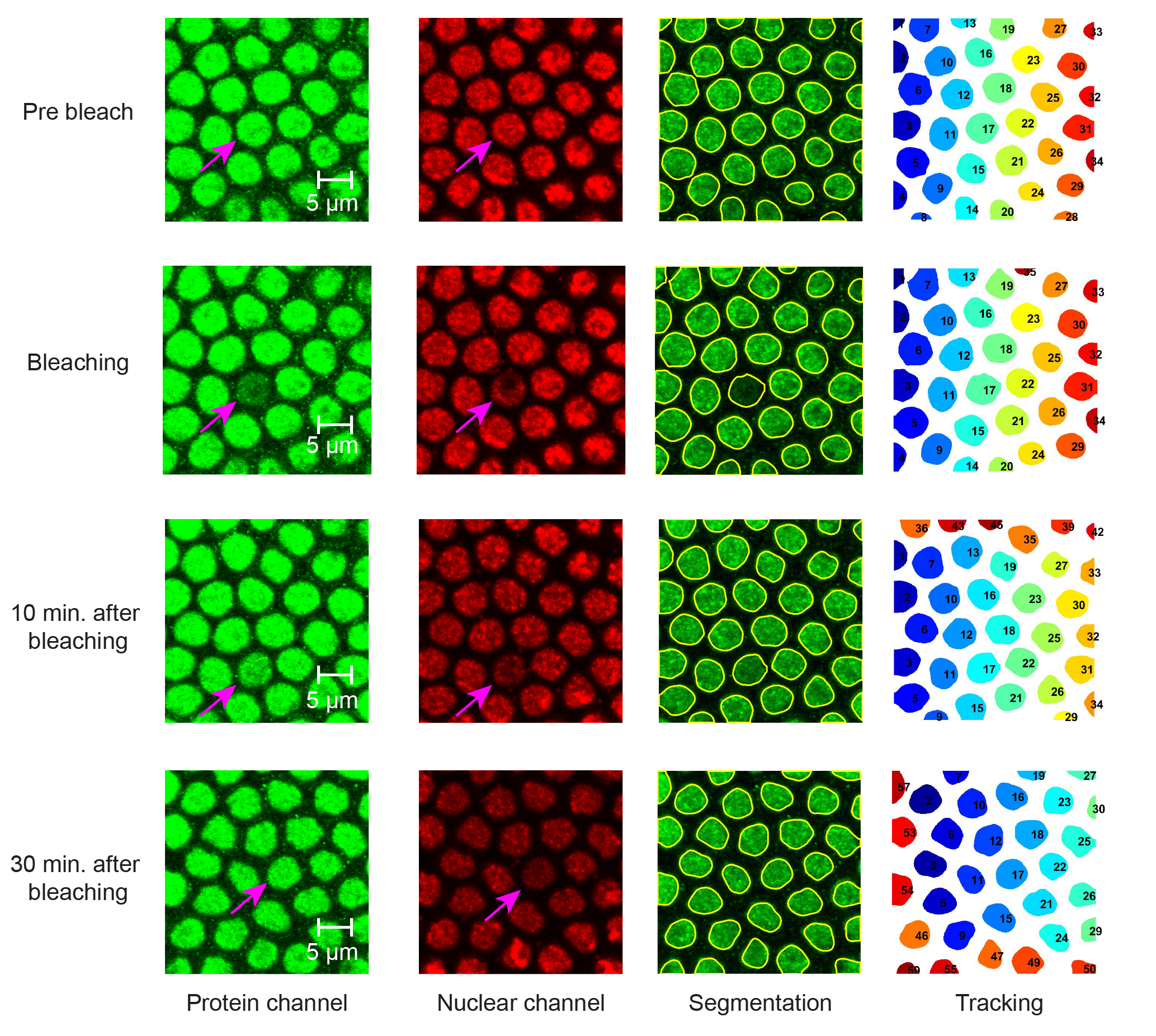}%
\caption{Results of image analysis when the nuclear int. $>>$ cytoplasmic intensity. The bleached nucleus (label number 17) was segmented and tracked through all the time frames along with its neighbors. }
\label{fig: image analysis nuc high cyt low}
\end{figure*}

\begin{figure*}[!t]
\centering
\includegraphics{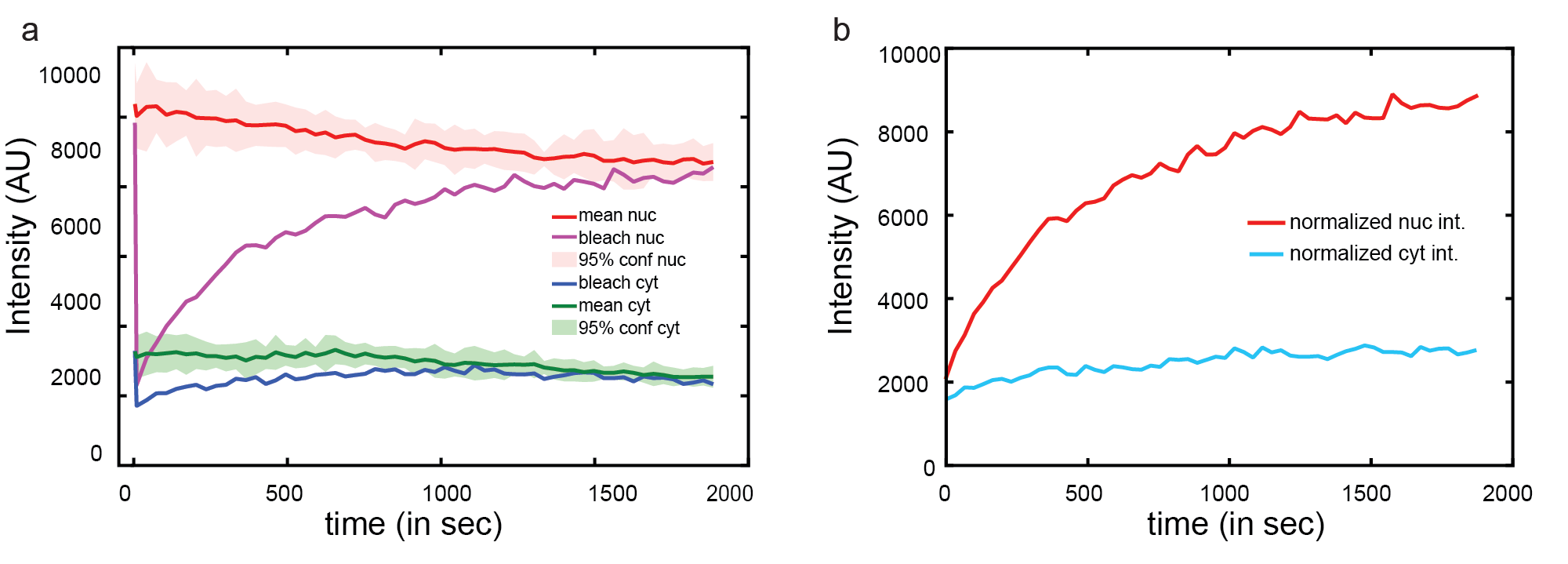}%
\caption{Plots of intensities extracted from nuclear protein channel for nuclei and cytoplasmic compartments. (a) Intensities of bleached nucleus and cytoplasm and neighboring nuclei and cytoplasmic compartments. (b) Intensities of the bleached nucleus and cytoplasm normalized by the mean intensities of neighboring cytoplasm and nucleus, respectively.}
\label{fig: nuc intensities}
\end{figure*}
\begin{figure}[!t]
\centering
\includegraphics{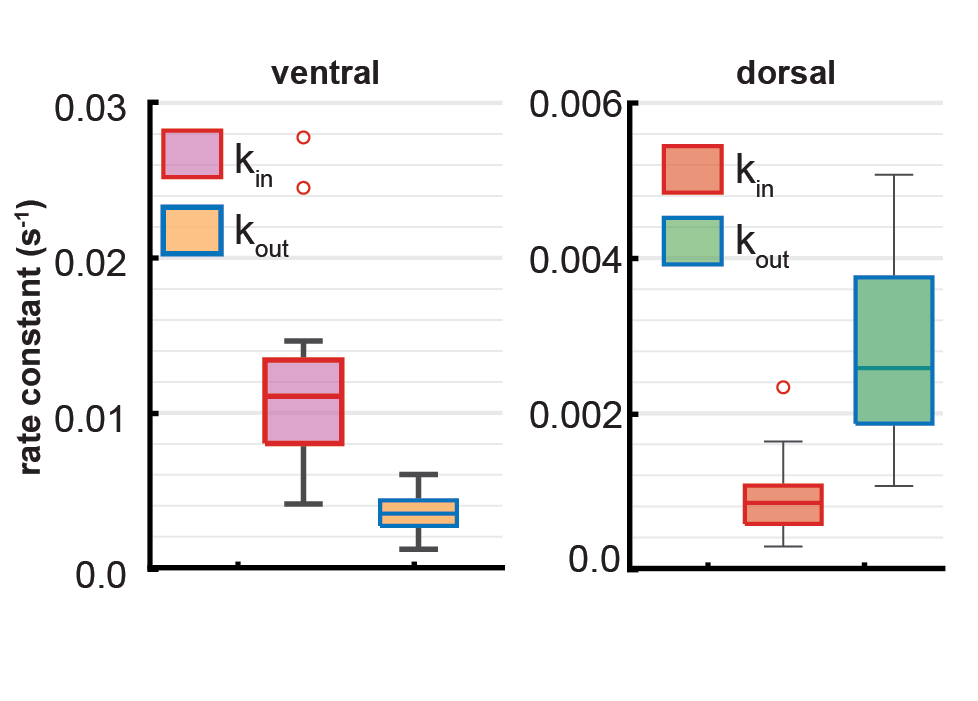}
\caption{Boxplots of the import and export rates from dorsal and ventral side.}
\label{fig:import_export_rates}
\end{figure}
\begin{figure*}[!t]
\centering
\includegraphics{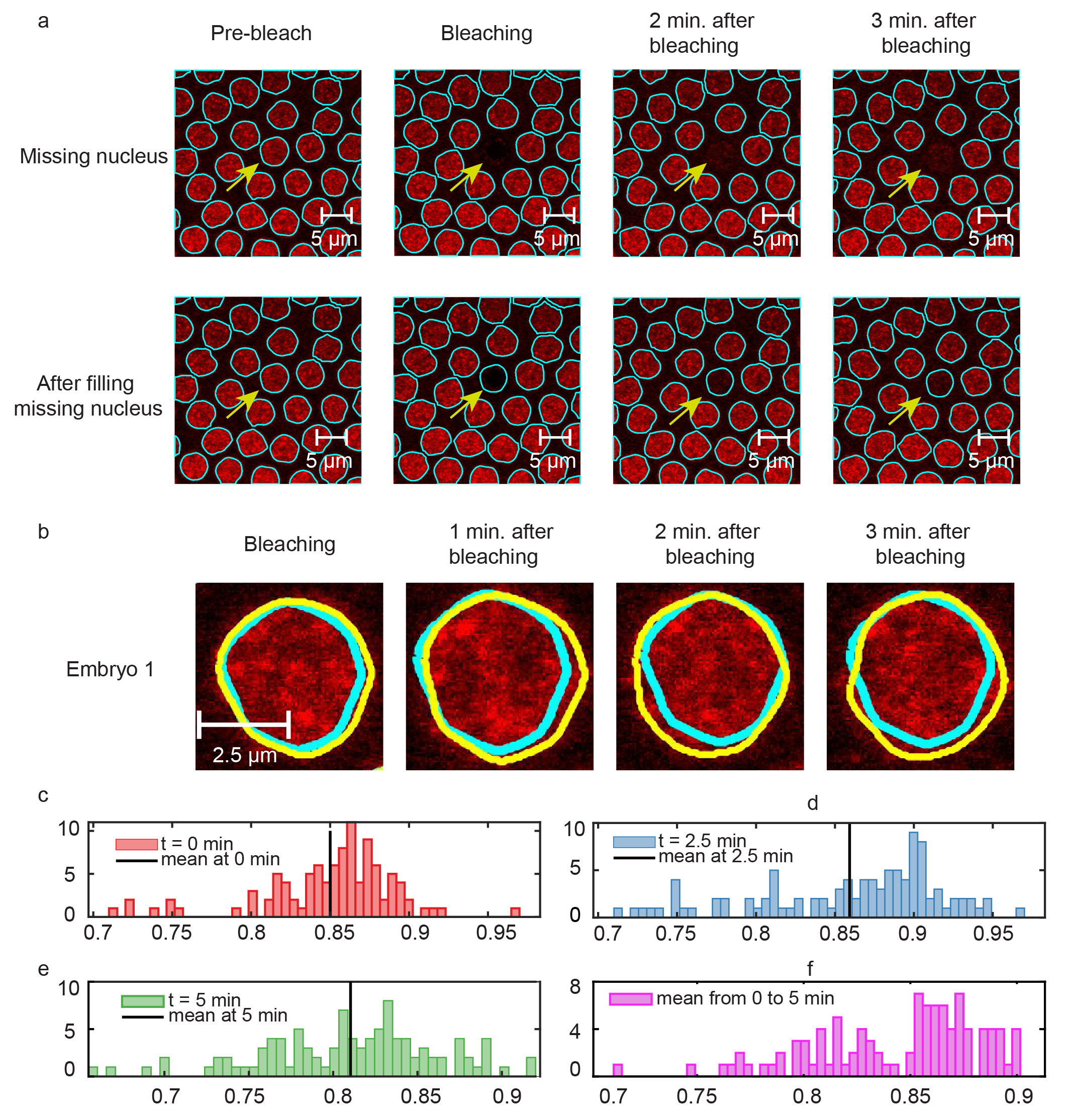}%
\caption{(a) Results of image analysis when bleached nucleus goes missing. We use the initial mask of the bleached nucleus as a reference and move it through the frames using the centroid locations of neighbors to generate the projected location of the missing nucleus. (b) Comparisons of masks obtained via segmentation and the masks obtained via shifting the mask from the initial frame using the centroid locations of neighbors to generate the projected location. (c)-(f) Histograms of IOUs comparing the masks from segmentation and masks shifted using neighboring locations.}
\label{fig: filling missing nuc ious}
\end{figure*}

We analyzed FRAP experiments from blastoderm-stage (1-3 h old) \textit{Drosophila} embryos expressing Dl-GFP and Histone-RFP (to mark the nuclei) \cite{carrell2017facilitated}. In these experiments, we took a snapshot of the region of interest (“Pre-bleach” in Fig. 2c), bleached a single nucleus near the center of the image at time t=0 (“Post bleach” in Fig. 2c), then imaged the nuclei in the region of interest for roughly 30 minutes (“Recovery” in Fig. 2c). 

\subsection{Segmentation}

In the segmentation phase (Fig. 5; see Methods for more details) of our study, we successfully delineated individual nuclei from the surrounding cytoplasm and other cellular structures. Utilizing the watershed segmentation technique, we accurately identified nuclear boundaries even in densely populated images with closely packed nuclei.

Figure 7 displays the segmentation outcomes when data were collected from the dorsal side of the embryo. In this scenario, the nuclear intensity in the protein channel is significantly lower than the cytoplasmic intensity. Due to the high contrast between nuclear and cytoplasmic regions in both channels, either could be used for segmentation purposes. If the nuclear protein channel is chosen, the only adjustment required would be to invert its intensities. Initially, in Fig. 7, the nuclei appear regularly shaped during the pre-bleach phase but exhibit irregular shapes in later time frames (beyond 10 minutes post-bleaching). Our segmentation method successfully delineates these nuclei throughout their shape transformations. In Fig. 8, where the nuclear intensity is significantly higher than the cytoplasmic intensity, we employ the nuclear channel for effective segmentation. However, there are instances where the bleached nucleus also vanishes from the nuclear channel, as shown in Fig. 11a. We address this challenge using the approach outlined in the Materials and Methods section under "Missing bleached nucleus."

\subsection{Tracking}

Following segmentation, we tracked the movement of each identified nucleus over time. The need for consistent nucleus tracking, where we assign to each nucleus a label that persists throughout the time course so that the identity of each nucleus is maintained, cannot be overstated. Our tracking algorithm leveraged the nuclei's relative positions and coordinated motion patterns, which were described as forming a constellation within the cellular space (see Methods). In Figure 7, we successfully tracked the nuclei across various timeframes, even as they underwent changes in shape since the relative positions of the centroids didn't change much. In Fig. 8, despite challenges posed by nuclei entering and leaving the frame, our approach maintained accurate tracking by utilizing a lookback feature. This feature was instrumental in re-identifying nuclei that temporarily moved out of the observational frame and then re-entered, ensuring continuity and consistency in our longitudinal analysis.

Given the application to FRAP experiments, one challenge facing our nuclear segmentation and tracking algorithms is generating masks for the bleached nucleus, which sometimes vanishes from both channels post-bleaching. We used the approach described in the Materials and Methods section to overcome this challenge. As such, nucleus tracking also affects the segmentation of the bleached nucleus in the frames during which it has vanished (Fig. 11a). To test these assumptions, we conducted a comparative analysis between masks generated under normal conditions and those derived using the neighboring location-based method, employing Intersection over Union (IoU) as our evaluation metric. The comparative results indicate that our approach yields an average IoU of .85 across more than 100 nuclei and 18 embryos, suggesting a high degree of accuracy (Fig. 11b-f). This performance metric further improves when the size of the nucleus remains consistent over the observed period.

\subsection{Extracting intensities of nuclei and cytoplasm}

The nucleus tracking algorithm described in the Materials and Methods section allowed us to accurately and consistently identify the fluorescence recovery of the bleached nucleus (Fig. 9a). Furthermore, our approach also extracts the intensities of the surrounding nuclei and cytoplasmic regions (Fig. 9a). These intensities of the surrounding nuclei and cytoplasm (Fig. 9a) show a slight dip which would indicate the occurrence of unintentional bleaching. We normalized the intensity of the bleached nucleus and its associated cytoplasm (Fig 9b) using 10 of its closest neighbors. The resulting normalized intensities (Fig. 9b) ensure that the measurements reflect true biological variations rather than artifacts of the imaging process.

\subsection{Estimating import and export rates}

The final analytical phase involved using the normalized fluorescence intensity data (Fig. 9b)  to estimate protein import and export rates. By fitting these intensities using the procedure described in the Methods section, we determined the rates at which the protein entered and exited the nucleus. Given the biological variation among the embryos, the boxplot of the import and export rates for Dl at two locations (dorsal and ventral) are presented (Fig. 10). The results show the variation in the rates based on the location which indicates that the spatial asymmetry of the Dl concentration.

Our work contributes an effective pipeline for segmenting and tracking nuclei, tailored specifically for biological data where the measurement of biophysical parameters from experimental image data is inherently challenging yet fundamentally important. This pipeline proves valuable in contexts requiring accurate identification and continuous monitoring of nuclei to decipher cellular behaviors and interactions.

\section{Conclusion}

Our study on FRAP experiments with blastoderm-stage \textit{Drosophila} embryos has yielded a robust image analysis pipeline for nuclear segmentation and tracking, demonstrating high adaptability and accuracy across varying conditions. This pipeline, crucial for identifying fluorescence recovery in bleached nuclei, employs a points registration algorithm to ensure consistent nucleus identification over time. A notable contribution of our work is the innovative strategy developed to track nuclei that vanish post-bleaching. In instances where the bleached nucleus becomes invisible in both nuclear and protein channels, we utilized the positional data of adjacent nuclei to infer the location of the missing nucleus. This approach offers a reliable solution to a common challenge in fluorescence recovery experiments. By accurately tracing the bleached nucleus across frames where it is otherwise undetectable, our method enhances the fidelity of nuclear tracking and segmentation.

Looking forward, the potential for automating segmentation tasks using generated masks as training data opens up new avenues for research, promising to refine the application of our pipeline in biological studies.

\newpage
\bibliographystyle{IEEEtran}
\bibliography{IEEEabrv, references}

\end{document}